\documentclass[aps,prd,reprint,superscriptaddress,nofootinbib,amsfonts,amssymb,amsmath,twocolumn,eprint]{revtex4-2}
\usepackage[utf8x]{inputenc}
\usepackage[dvipdf,dvips,dvipdfmx]{graphicx}
\usepackage{float}
\usepackage{wrapfig}
\usepackage{bm}
\usepackage{color}
\usepackage{comment}
\usepackage{xcolor}
\usepackage[normalem]{ulem}
\usepackage{hyperref}
\hypersetup{
colorlinks=true,
citecolor=blue,
citebordercolor=red,
linktoc=all,
linkcolor=blue,
urlcolor=blue
}
\usepackage{subfigure}
\graphicspath{{Figures/}}

\usepackage[normalem]{ulem}

\begin{document}


\title{
Asymptotic safety meets tensor field theory: towards a new class of gravity-matter systems
}%

\author{Astrid Eichhorn}
    \email{eichhorn@thphys.uni-heidelberg.de}
 \affiliation{Institute for Theoretical Physics, Heidelberg University, Philosophenweg 12, 16 and 19, 69120 Heidelberg, Germany}
\author{Razvan Gurau}%
  \email{gurau@thphys.uni-heidelberg.de}
 \affiliation{Institute for Theoretical Physics, Heidelberg University, Philosophenweg 12, 16 and 19, 69120 Heidelberg, Germany}
 \author{Zois Gyftopoulos}
    \email{gyftopoulos@thphys.uni-heidelberg.de}
 \affiliation{Institute for Theoretical Physics, Heidelberg University, Philosophenweg 12, 16 and 19, 69120 Heidelberg, Germany}

\begin{abstract}

Combining asymptotically safe quantum gravity with a tensor field theory, 
we exhibit the first example of a theory with gravity and scalar fields in four dimensions which may realize asymptotic safety at a non-vanishing value of the scalar quartic coupling. 

We first present (further) evidence that in the asymptotic-safety paradigm, quantum fluctuations of gravity generically screen the quartic couplings in (multi-)scalar models. For a tensor field theory in which the scalar field transforms under an internal $O(N)^3$ symmetry, this has the effect of replacing asymptotic freedom, recently discovered at large $N$ on a fixed flat background,
by an interacting fixed point in the presence of quantum gravity. The fixed point originates from the competition between the 
effects of the matter self-interactions which, contrary to the usual scalar models, are antiscreening, and the screening gravitational effects.
\end{abstract}

\maketitle

\section{\label{sec:level1}Introduction}
Quantum field theories can exist as effective theories with a finite range of scales over which they provide an accurate description of a physical system, or as fundamental theories, which are self-consistent over an infinite range of scales. In four dimensions, scalar $\phi^4$ theory belongs to the first category, because it becomes trivial if one attempts to extend its range of validity to all scales \cite{Frohlich:1982tw,Luscher:1987ay,aizenman2021marginal}. One may try to circumvent this in various ways. A partial list of ideas includes 
i) introducing a large internal symmetry group \cite{Berges:2023rqa}, ii) coupling the theory to gravity \cite{Narain:2009fy, Eichhorn:2017als},
iii) adding derivative interactions 
\cite{Buccio:2022egr} iv) coupling to a gauge-Yukawa sector \cite{Litim:2014uca} v) changing the spacetime dimensionality \cite{Litim:2018pxe} vi) making spacetime non-commutative \cite{Grosse:2004by,Sfondrini:2010zm}
vii) continuing to negative coupling \cite{Romatschke:2023fax}, and the list goes on. 

In the present paper, we combine
the first two ideas and introduce an internal $O(N)^3$ symmetry together with a coupling to gravity. Our motivation for doing so is twofold. First, in order to advance our understanding of quantum field theories and their ultraviolet (UV) 
structure, we study how a theory that is
UV complete due to a Renormalization Group (RG) fixed point changes
i.e., how its
RG flow and fixed-point structure is affected, when
gravity fluctuations are taken into account. 
Second, in the asymptotic-safety scenario for quantum gravity one will eventually need to introduce new matter sectors beyond the Standard Model in order to account for dark matter, neutrino masses, and matter-antimatter asymmetry. Additional scalars are popular candidates for the first and third of these challenges.
In order to develop a comprehensive picture of possible building blocks for these sectors, one must understand the interplay between quantum gravity and non-conventional scalar field theories, such as the one introduced in \cite{Berges:2023rqa}.

\paragraph{The asymptotic-safety scenario for gravity.} In order to set the stage, we briefly review the main ideas of the asymptotic-safety scenario for gravity \cite{Hawking:1979ig}, and the current understanding of the interplay of asymptotically safe gravity with matter fields.

In an asymptotically safe theory the microscopic (UV) physics is governed by an interacting RG fixed point. For gravity, such a UV fixed point can in principle remove the apparent lack of predictivity arising from perturbative non-renormalizability: the infrared (IR) theory is uniquely determined by the initial conditions along the (putatively few) IR repulsive, or relevant, directions, while the (infinitely many) non-renormalizable operators are just IR attractive, or irrelevant, directions of the fixed point. Initializing an RG trajectory in the deep UV close
to the finite dimensional critical (hyper)surface spanned by the relevant directions yields infinitely many relations between the running couplings in the low energy effective field theory, thus restoring predictivity.

The existence of an interacting RG fixed point can be checked, and the dimensionality of its critical (hyper)surface be calculated,  
within the functional RG framework (FRG), introduced in \cite{Wetterich:1992yh, Morris:1993qb} and adapted to gravity in \cite{Reuter:1996cp}, by testing the response of the couplings of the theory to the variation of an \emph{unphysical} auxiliary scale (the RG scale) $k$ that tracks which quantum fluctuations have been integrated over. 

This RG scale is not a physical momentum scale. The correct physical UV behavior of the theory is encoded in the effective action at $k \rightarrow 0$, but evaluated at high momentum or high curvature scales. In other words, 
after identifying a fixed point using the RG scale  $k$, one needs to check that the momentum-dependent vertex functions are not singular. 
As the running with respect to $k$ and the running with respect to physical momentum can be different, see, e.g., \cite{Buccio:2023lzo}, this is a non-trivial check. This point has in the past led to some discussion \cite{Donoghue:2019clr,Bonanno:2020bil}.
In the modern incarnation of the asymptotic-safety scenario, one calculates the momentum-dependent vertex functions (or form-factors) in the presence of the RG scale $k$ and tests whether in the physical limit $k \rightarrow 0$, the momentum-dependence yields finite observables, see \cite{Knorr:2019atm,Pawlowski:2020qer,Knorr:2022dsx} for reviews of such calculations and \cite{Knorr:2021iwv,Knorr:2022lzn,Pastor-Gutierrez:2024sbt} for recent results on the physical scale dependence. These results are based on a large body of previous work which provides compelling evidence for an asymptotically safe fixed point, see  \cite{Alkofer:2020vtb,Platania:2020knd,Falls:2020qhj,deBrito:2020dta,Bonanno:2021squ,Laporte:2021kyp,Ferrero:2021xqg,Baldazzi:2021fye,Basile:2021krr,Ohta:2021bkc,Daas:2021abx,Sen:2021ffc,Knorr:2021niv,Fehre:2021eob,Baldazzi:2021orb,deBrito:2021pyi,Eichhorn:2021tsx,Eichhorn:2021qet,deBrito:2021akp,Sen:2022xlp,Knorr:2022ilz,deBrito:2022vbr,Eichhorn:2022ngh,Wetterich:2022bha,Pastor-Gutierrez:2022nki,Eichhorn:2022vgp,Wetterich:2022ncl,Ferrero:2022hor,Ferrero:2022dpk,deBrito:2023kow,deBrito:2023myf,Eichhorn:2023jyr,deBrito:2023ydd,Knorr:2023usb,Saueressig:2023tfy,Baldazzi:2023pep,Eichhorn:2023gat,Becker:2024tuw,Saueressig:2024ojx,Wetterich:2024ieb,Korver:2024sam,Eichhorn:2024wba,Knorr:2024yiu,Falls:2024noj,Brenner:2024bps,Ferrero:2024rvi,Saueressig:2025ypi} for the most recent developments, \cite{Souma:1999at,Reuter:2001ag,Lauscher:2001ya,Litim:2003vp,Codello:2008vh,Benedetti:2009rx,Manrique:2011jc,Dona:2013qba,Falls:2013bv} for key earlier results, \cite{Eichhorn:2018yfc,Bonanno:2020bil,Eichhorn:2022jqj,Eichhorn:2022gku,Eichhorn:2023xee,Pawlowski:2023gym,Saueressig:2023irs} for reviews and \cite{Reuter:2019byg,Reichert:2020mja,Eichhorn:2020mte,Basile:2024oms} for books and lecture notes.

Concerning the addition of matter fields, studies to date indicate that asymptotically safe gravity  
extends to some settings with matter, \cite{Dona:2013qba,Meibohm:2015twa,Biemans:2017zca,Christiansen:2017cxa,Alkofer:2018fxj,Wetterich:2019zdo, Korver:2024sam} and specifically scalar matter \cite{Narain:2009fy, Dona:2013qba, Percacci:2015wwa, Labus:2015ska, Meibohm:2015twa, Dona:2015tnf, Biemans:2017zca, Alkofer:2018fxj, Eichhorn:2018akn,Burger:2019upn,Wetterich:2019zdo, Laporte:2021kyp, Ohta:2021bkc,Sen:2021ffc,deBrito:2021pyi,deBrito:2023myf}, see also the review \cite{Eichhorn:2022gku},  and that such  gravity-matter models impose additional constraints on the matter content (e.g., some free parameters of the Standard Model become calculable from first principles  \cite{Shaposhnikov:2009pv,Eichhorn:2017ylw,Eichhorn:2017lry, Eichhorn:2018whv}). 
 However, if one considers only the scalar sector coupled to asymptotically safe gravity, the UV behavior of the matter sector is not improved. In fact the gravity contribution is ``screening" in this case, i.e., gravity fluctuations shift the UV Landau pole of the matter coupling to a lower RG scale. From a phenomenological perspective, this implies that the quartic coupling of the Standard Model Higgs field must vanish at the Planck scale and arise below the Planck scale as a radiative correction due to the gauge fields and fermions (most importantly the top quark). This yields a prediction of the ratio of Higgs mass to electroweak scale \cite{Shaposhnikov:2009pv,Eichhorn:2021tsx,Pastor-Gutierrez:2022nki}, which is rather close to the experimental result.

We will revisit  the UV fate of the matter sector in the context of a large number of scalar fields below.

 \paragraph{Tensor field theories.} In a parallel, and so far largely unrelated (see, however \cite{Eichhorn:2019hsa}), research direction, random tensors have been introduced a long time ago \cite{Ambjorn:1990ge,Sasakura:1990fs} with the aim of reproducing the success of random matrices as theories of random surfaces \cite{DiFrancesco:1993cyw} 
in higher dimensions.
However, generalizing the $1/N$ expansion of matrices \cite{tHooft:1973alw} to tensors proved 
non-trivial \cite{Gurau:2011xq,guruau2017random}. It turns out that tensors exhibit a new large-$N$ limit, the so called \emph{melonic} limit \cite{Bonzom:2011zz}, which is 
in fact simpler than its matrix counterpart. Building on the melonic limit, tensor quantum mechanical models have been introduced and related to the Sachdev–Ye–Kitaev model \cite{Sachdev:1992fk,Kitaev}, but without disorder \cite{Witten:2016iux}, providing a first link between tensor models and quantum gravity via holography \cite{Narovlansky:2023lfz}.
Subsequently, tensor field theories \cite{Giombi:2017dtl,Bulycheva:2017ilt} have been extensively studied as possible conformal field theory duals of AdS gravity theories in higher dimensions.

The prototypical tensor field theory model is the quartic $O(N)^3$ model 
\cite{Carrozza:2015adg,Giombi:2017dtl,Benedetti:2019eyl}, which has been extensively studied, both in its short and in its long-range versions \cite{Giombi:2017dtl,Benedetti:2019eyl,Benedetti:2019ikb,Benedetti:2021wzt,Benedetti:2020sye}. Notably, in less than $4$ dimensions, the $O(N)^3$ model proved instrumental in identifying a new class of conformal field theories, the so called \emph{melonic} CFTs, which
are reached when one of the couplings (dubbed \emph{tetrahedral}, see below) is chosen imaginary \cite{Gurau:2019qag}. In spite of the presence of an imaginary coupling, all indications to date point to the fact that the $N\to \infty$ theory is in fact a unitary conformal field theory \cite{Benedetti:2019ikb}.  

The $O(N)^3$ model has recently been studied also in dimension $d=4$ with surprising conclusions: sticking to an imaginary tetrahedral coupling, one obtains an \emph{asymptotically free} and \emph{stable} model \cite{Berges:2023rqa,Berges:2024ydj}. This is in stark contrast with a "vanilla" scalar field theory with quartic interaction, which is either trivial or ill-defined.
The imaginary coupling breaks unitarity; however,
in the strict large-$N$ limit no unitarity breaking effect is present (i.e., the quantum effective action is real). 

\paragraph{Bringing the two together.} In this paper we study the interplay between the quantum fluctuations of gravity and the $O(N)^3$ tensor field theory model. The fundamental question we aim to clarify is the following: what are the quantum-gravity-induced effects on the UV
behavior of the model? More precisely, after taking into account quantum-gravity fluctuations, is asymptotic freedom preserved, replaced by a UV fixed point, or simply lost?

\section{Gravitational contributions to the flow of quartic couplings}

\subsection{Principled-parameterized approach to asymptotically safe gravity-matter systems}

We work in the framework of the \emph{principled-pa\-ra\-me\-terized} approach  \cite{Eichhorn:2018whv} to parameterize the effect of new physics in the beta functions. 
In the most general setting, the $\beta_{\lambda_c}$ function of a matter coupling  $\lambda_c$ takes the form\footnote{Here $t = \ln (k/k_0)$ with $k$ the RG scale and $k_0$ a reference scale.} 
\begin{equation}
\beta_{\lambda_c} = \partial_t \lambda_c = (\Delta_c - d)\lambda_c + \beta^n_{\lambda_c}  - f_c \lambda_c + \dots \;,
\end{equation}
where 
$\Delta_c$ is the canonical dimension of the corresponding operator, $\beta^n_{\lambda_c}$ encode the quantum correction due to the matter self interactions, $f_c \lambda_c$ is the leading quantum correction due to gravitational fluctuations, and the ellipses stand for the higher order contributions of the  gravitational fluctuations. In the principled-parametrized framework, the first order gravitational contributions to $\beta_c$, encoded in $f_c$, is constrained by general principles.
First, because gravity couples to all the terms in the Lagrangian of the matter field, gravity fluctuations back react on all the matter couplings at linear order in $\lambda_c$, so that the coefficient $f_c$ only depends on the gravitational couplings.
Second, as gravity is ``blind" to internal symmetries, operators built from different invariants of the internal symmetry group are indistinguishable from the point of view of spacetime symmetries. Thus, $f_c$ is the same for all the couplings that correspond to operators with the same tensor structure under Euclidean rotations.\footnote{For example $\phi^{20}$ and $\phi^{10} \phi^{10}$ have the same tensor structure under Euclidean rotations but are different invariants under internal symmetries hence get the same $f_{\lambda}$, while $\phi^{10}\Box\phi^{10}$ and $\phi^{18}$ have the same mass dimension but receive different gravitational contributions to their beta functions.} 
Consequently, there is a single $f_g$ for all gauge couplings $g$ irrespective of whether, for instance, the gauge group is Abelian or not,\footnote{In the case of $U(1)$  the gauge coupling is read off from the gauge field / matter fields interaction terms, as there is no self interaction of the Abelian gauge field.} a single $f_{\lambda}$ for all quartic couplings $\lambda$, and so on. Third, in the gravitational fixed-point regime, which we assume to be the transplanckian regime
$k \geq M_{\rm Planck}$, the $f_\lambda$ is effectively constant as the gravitational couplings go to their fixed point values:
\begin{eqnarray}
         f_{\lambda}(G_k ,\Lambda_k,\dots) \xrightarrow{k \gg M_{pl}} f_{\lambda}(
         G_{\star},\Lambda_{\star},\dots)< \infty \;,
\end{eqnarray}
where $\Lambda_k$ and $G_k$ are the running dimensionless cosmological constant and Newton coupling\footnote{The ellipses denote the dimensionless gravitational couplings corresponding to operators of higher mass dimension}.
Below the Planck scale, we depart from the fixed point regime and $f_{\lambda}\left(G_k,\Lambda_k\right)$ runs quickly to zero as $G_k \sim k^2 /M_{\rm Planck}$, leading to a subplanckian flow of $\lambda$ in which gravity decouples. 

This framework, initially motivated by asymptotic safety, can provide a UV completion for various sectors of the Standard Model (SM). This, for instance, works if $f_c>0$, that is
the gravitational contribution is antiscreening. Not only can gravity contributions potentially improve the UV behavior of the matter couplings, but also in this framework some free parameters of the SM are constrained. For instance, the low-energy values of the SM couplings are constrained by imposing that the infrared theory sits on the critical surface. The principled-parametrized framework has been used to explore whether asymptotic safety  
can be realized in the presence of matter and whether predictions for couplings arise in subsectors of the SM \cite{Eichhorn:2018whv,Alkofer:2020vtb}, in dark-matter models \cite{Reichert:2019car,deBrito:2023ydd}, in models for neutrino masses \cite{Held:2019vmi,Eichhorn:2022vgp,Kowalska:2022ypk,Chikkaballi:2023cce} and other BSM settings \cite{Eichhorn:2017muy,Eichhorn:2019dhg,Kowalska:2020gie,Kowalska:2020zve,Chikkaballi:2022urc}. Notably, the gravity contribution is antiscreening in the $U(1)$ sector \cite{Daum:2009dn,Harst:2011zx,Folkerts:2011jz,Christiansen:2017cxa} which 
suggests that when accounting for gravity the $U(1)$ Landau pole is traded for an interacting fixed point \cite{Harst:2011zx, Eichhorn:2017lry,Christiansen:2017gtg,deBrito:2022vbr}. However, as mentioned in the introduction, this is not the case for the scalar sector, for which gravity has been found to be screening, i.e., $f_{\lambda} < 0$ \cite{Narain:2009fy, Percacci:2015wwa, Labus:2015ska,Eichhorn:2017als}.

In this paper, we apply such considerations to a scalar field described by a $O(N)^3$ quartic tensor model. As the quartic couplings are marginal in $d=4$ dimensions, their $\beta$ functions read:
\begin{eqnarray}
    \partial_t \lambda_i = - f_{\lambda} \lambda_i + \beta^{n}_{\lambda_i} + \dots \label{eq:paramUVcompletion} \; ,
\end{eqnarray}
and for the matter self interaction quantum corrections $ \beta^{n}_{\lambda_i}  $ we will use the lowest non trivial order, which turns out to be the 2-loop order \cite{Zinn-Justin:1989rgp,Giombi:2017dtl}.

\subsection{Calculation of 
\texorpdfstring{$f_{\lambda}$}{flambda} by functional Renormalization Group techniques}

In this section we calculate the gravitational contribution
$f_{\lambda}(G,\Lambda)$ from first-principles and in a simple approximation. This serves a two-fold purpose: first, we review these type of calculations for readers not familiar with FRG calculations. Second, we supplement existing results in the literature by working in an unusual choice of gauge parameters for the gravitational fluctuations. Studying the gauge dependence of physical quantities (e.g., critical exponents) provides a measure of the quality of our approximation, because the gauge dependence should vanish entirely once no approximation is made in the calculation. 

Let us consider a field theory for a (Lorentz) scalar field:
\begin{eqnarray}
    \phi_{\textbf{a}} \equiv \phi_{a_1 \dots a_D}\;\;,\;\; a_i = 1,\dots,N  \label{eq:tensorfielddefinition} \;,
\end{eqnarray}
having $\prod_{i=1}^D N_i$ components. We call $i$ the color of the index $a_i$. 
The tensor $\phi_{\textbf{a}} $ has no symmetry properties under permutations of its indices and transforms in the external 
tensor
product of fundamental representations of the orthogonal group:
\begin{eqnarray}\label{eq:globalsymmgroup}
    O(N_1)\times \dots \times O(N_D) \;  ,
\end{eqnarray} 
that is 
\begin{eqnarray}
   \phi'_{a^1,\dots,a^D} = \sum_{b^1,\dots,b^D} O^{(1)}_{a^1b^1}\cdots O^{(D)}_{a^Db^D} \phi_{b^1,\dots,b^D} \;. 
\end{eqnarray}
While below we will choose all the $N_i$'s equal, everything goes through for indices of different ranges.

We study the gravitational contribution to the multi-scalar quartic interaction $\lambda_k^{\textbf{abcd}}$ 
using Functional Renormalization Group (FRG) methods \cite{Wetterich:1992yh,Morris:1993qb,Reuter:1996cp}. We introduce an IR-regulator with a choice of shape function generally known as Litim regulator \cite{Litim:2001up}:
\begin{equation}
    \mathcal{R}_k(p) = \mathcal{Z}_k k^2 \left(1 - \frac{p^2}{k^2}\right)\Theta\left(1 - \frac{p^2}{k^2}\right) \;,
\end{equation}
that suppresses Fourier modes of the field with mo\-men\-tum-squared $p^2$
below $k^2$, properly adjusted with wave-function renormalization factors and tensor structures (encoded in $\mathcal{Z}_k$).

The flowing action, that is the $k$-dependent functional $\Gamma_k$, interpolates between the microscopic (bare) action $\Gamma_{k\to\infty} \equiv S_{\Lambda}$ and the quantum effective action $\Gamma_0 \equiv \Gamma$, and satisfies the (formally) exact Wetterich equation \cite{Wetterich:1992yh,Morris:1993qb}:
\begin{eqnarray}\label{eq:Wett}
   \partial_t \Gamma_k = \frac{1}{2}{\rm Tr}\left\{(\Gamma_k^{(2)}+ \mathcal{R}_k)^{-1}(\partial_t \mathcal{R}_k)\right\} \; .
\end{eqnarray}
As the regulator $R_k(p)$ suppresses  the low-momentum modes and $\partial_t R_k$ vanishes for high momenta, the main contribution to the right-hand side of Eq.~\eqref{eq:Wett} comes from loop momenta close to $k$. This realizes a Wilson-type,  momentum-shell integration of the underlying path integral, in which the main change of the effective dynamics at $k$ is driven by quantum fluctuations with momentum $p^2 \approx k^2$.
While the Wetterich equation is exact, as $(\Gamma_k^{(2)}+ \mathcal{R}_k)^{-1}$ carries the full field and momentum dependence, it nevertheless has a one-loop structure which makes it very efficient for practical calculations in a broad range of fields, from statistical physics and strongly-correlated fermion systems to Quantum Chromodynamics, the Standard Model and quantum gravity, see 
\cite{Dupuis:2020fhh} for a review.
The scale dependence of any coupling is extracted by projecting the right-hand-side onto the corresponding field monomial, and it is at this stage that one needs to use a systematic expansion and truncate $\Gamma_k$.

When applying this equation to gravity, one uses the background field method which consists in 
splitting the metric into a fixed background $\bar{g}_{\mu\nu}$ and a fluctuation $h_{\mu\nu}$, $ g_{\mu\nu} = \bar{g}_{\mu\nu} + h_{\mu\nu}$. For the matter part of the effective action, we choose a truncation ansatz with vanishing scalar mass and minimal coupling, such that the flowing action reads:
\begin{eqnarray}
\Gamma_k&=& \frac{\mathcal{Z}_{S}}{2}\int d^4x\sqrt{g} g^{\mu\nu}\partial_{\mu}\phi_{\textbf{a}}\partial_{\nu}\phi_{\textbf{a}}\nonumber\\
&{}&+ \frac{\lambda_k^{\textbf{abcd}}}{4!}\mathcal{Z}_S^2\int d^4x \sqrt{g}\phi_{\textbf{a}}\phi_{\textbf{b}}\phi_{\textbf{c}}\phi_{\textbf{d}} \; , 
\end{eqnarray}
where we account for the $k$-dependence of the fields via a field-independent wave-function renormalization 
$ \phi_{\textbf{a}} \rightarrow \sqrt{\mathcal{Z}_S} \cdot \phi_{\textbf{a}}$.
 In this choice of truncation, we rely on a set of previous results that indicate a \emph{near-perturbative} character of the asymptotically safe fixed point. Such a near-perturbative character means that most critical exponents are only slightly shifted compared to their perturbative values, see \cite{Falls:2013bv, Falls:2014tra,Falls:2017lst,Falls:2018ylp, Eichhorn:2020sbo}. The main exception is the critical exponent for the Newton coupling, which must switch sign for an asymptotically safe fixed point to be realized.
Further indications for near-perturbativity also arise from the study of symmetry-identities, which are well-compatible with a relatively weakly coupled fixed point \cite{Eichhorn:2018ydy}.

While we are ultimately interested in particular choices for $\lambda_k^{\textbf{abcd}}$ corresponding to  specific
contraction patterns for the indices, we work first with a general $\lambda_k^{\textbf{abcd}}$ and check that the lowest order gravitational contribution to $\beta_{\lambda_k^{\textbf{abcd}}}$ is indeed $- f_{\lambda} \lambda_k^{\textbf{abcd}}$, hence respects the symmetry of $\lambda_k^{\textbf{abcd}}$. This calculation confirms in this explicit example that gravity is ``blind" to internal symmetries and gravitational corrections do not violate global symmetries in the near-perturbative, Euclidean regime that our calculations explore. 

\begin{figure}[t!]
    \includegraphics[width=0.45\linewidth]{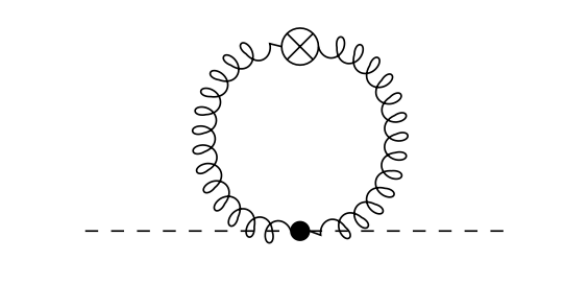}\quad
    \includegraphics[width=0.45\linewidth]{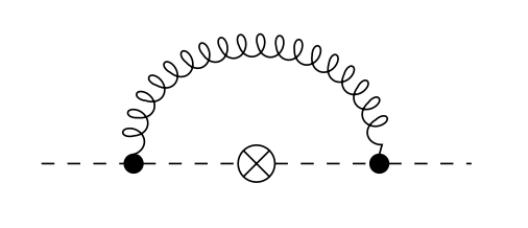}
    \caption{\label{fig:anomdimgraphs}Diagrams contributing to the anomalous dimension. Dashed lines denote the scalar field, curly lines the metric and the crossed circle denotes a regulator insertion. The two-vertex-diagram occurs in two versions, with the regulator inserted either on the dashed (as displayed) or curly line (not shown).}
\end{figure}

The running of the wavefunction renormalization $\mathcal{Z}_S$ defines the
anomalous dimension $ \eta_S = - \partial_t \left(\ln \mathcal{Z}_S\right) $
of the tensor field $\phi_\textbf{a}$.
A gravitational contribution to the anomalous dimension is present at one loop, whereas matter fluctuations start contributing at the two-loop order.
The flow of $\eta_S$ is calculated by evaluating the right-hand side of the flow equation for the simplest possible choice of background metric, that is flat\footnote{One can convince oneself, using, e.g., heat-kernel techniques, that the choice of a non-flat metric does not change the result and simply makes the evaluation of the trace in the Wetterich equation slightly more involved.},
leading to:
\begin{eqnarray}\label{eq:projectionprescription} 
     \eta_S = -\frac{2}{\mathcal{Z}_S} \partial_{p^2}\left(\frac{\delta^2}{\delta \phi_{\textbf{a}}(p)\delta \phi_{\textbf{a}}(-p)}\left.\partial_t \Gamma_k\right|_{\phi_\textbf{a} = 0}\right)\;,
\end{eqnarray}
which extracts the $p^2$-dependent part of the 1-loop self-energy diagrams depicted in Fig.~\ref{fig:anomdimgraphs}. This generically leads to an algebraic equation for the anomalous dimension $\eta_S$ of the form:
\begin{eqnarray}
    \eta_S = \left(\alpha(\Lambda_k) + b(\Lambda_k)\eta_h +c(\Lambda_k)\eta_S \right)G_k \;,
\end{eqnarray}
where $\eta_h$ denotes the anomalous dimension of the gravitational fluctuation $h_{\mu\nu}$.

\begin{figure}
        \includegraphics[width=0.4\linewidth]{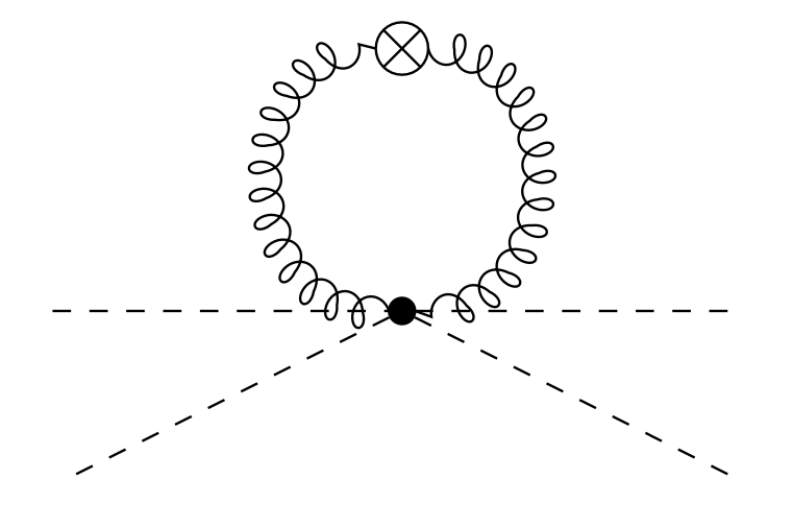}
        \includegraphics[width=0.4\linewidth]{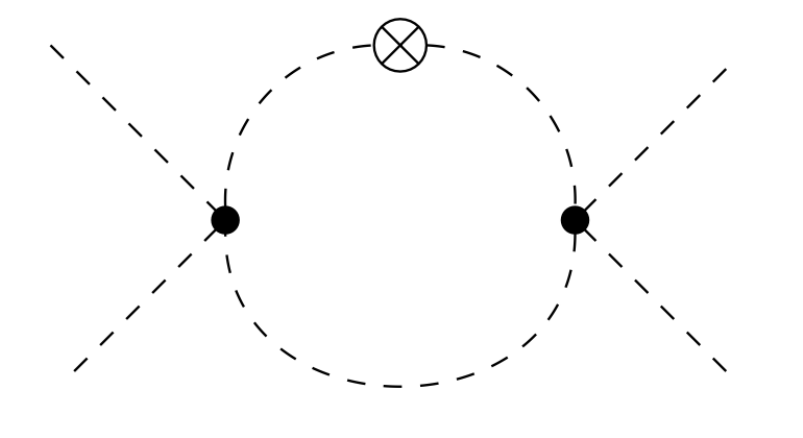}
    \caption{\label{fig:couplingFRGgraphs}4-point diagrams with non-trivial contribution at criticality. The vertices correspond to the fully dressed $\Gamma^{(6)}_{\phi\phi hh \phi\phi}$ and $\Gamma^{(4)}_{\phi\phi\phi\phi}$ respectively.}
\end{figure}
The flow of $\lambda_k^{\textbf{abcd}}$ is obtained by projecting Eq.~\eqref{eq:Wett} onto the quartic self-interaction term, after setting $\phi_{\textbf{a}}= \rm const$ to remove momentum-dependent self-interactions:
\begin{align}
    \partial_t \lambda_k^{\textbf{abcd}} &=  2\eta_S \lambda_k^{\textbf{abcd}} + \frac{4!}{\mathcal{Z}_S^2}\left.\left(\frac{\delta^4 (\partial_t \Gamma_k)}{\delta \phi_\textbf{a} 
    \phi_\textbf{b}
    \phi_\textbf{c}
    \phi_{\textbf{d}}
    } 
    \right)\right|_{\phi = 0} \; .
\end{align}
Its evaluation in the massless limit reduces to calculating the 1-loop diagrams in Fig.~\ref{fig:couplingFRGgraphs}.
In order to obtain the propagator of metric fluctuations $h_{\mu\nu}$, we need to fix the gauge and choose gauge parameters $\beta=1, \alpha=1$ in the gauge-fixing action $S_{\rm gf} = \frac{1}{32 \pi G_k k^{-2}}\frac{1}{\alpha}\int d^dx \sqrt{\bar{g}}F_{\mu}\bar{g}^{\mu\nu}F_{\nu}$ with $F_{\mu} = \left(\delta_{\mu}^{\lambda}\bar{D}^{\kappa} - \frac{1+\beta}{4} \bar{g}^{\kappa\lambda}\bar{D}_{\mu}\right)h_{\kappa \lambda}$. We obtain:
\begin{widetext}
    \begin{align}
    &\partial_t \lambda_k^{\bm{abcd}} 
    = (d-4)\lambda_k^{\bm{abcd}} + 2 \eta_S\lambda_k^{\bm{abcd}} +\frac{16 \pi G_k \cdot\lambda_k^{\bm{abcd}}}{(4\pi)^{\frac{d}{2}}\Gamma\left(\frac{d}{2}\right)}\frac{d+1}{d+2}\frac{(d+2 -\eta_h)}{(1-2\Lambda_k)^2} + \frac{12}{(4\pi)^{\frac{d}{2}}\Gamma\left(\frac{d}{2}\right)}\frac{(d+2 -\eta_S)}{d(d+2)}\lambda_k^{\bm{abef}}\lambda_k^{\bm{cdef}} \label{eq:quarticflow}
    ,\\ \nonumber\\
    & \eta_S = - \frac{32\pi G_k}{(4\pi)^{\frac{d}{2}}\Gamma\left(\frac{d}{2}\right)}\left[ \frac{2}{d+2}\frac{1}{(1-2\Lambda_k)^2}\left(1-\frac{\eta_h}{d+4}\right) + \frac{2}{d+2}\frac{1}{1-2\Lambda_k}\left(1-\frac{\eta_S}{d+4}\right) +\frac{(d+1)(d-4)}{2d(1-2\Lambda_k)^2}\left(1-\frac{\eta_h}{d+2}\right) \right].\label{eq:anomdimflow}
\end{align}
\end{widetext}

In the absence of gravity, $G_k,\Lambda_k \to 0$, and in $d=4$ dimensions, the flow we obtain matches the 1-loop result in dimensional regularization \cite{Zinn-Justin:1989rgp}. 
From Eq.~\eqref{eq:quarticflow}, we identify  the gravitational coefficient in $d=4$:
\begin{eqnarray} \label{eq:flambdaapprox}
     f_{\lambda}\left(G_k,\Lambda_{k}\right) = -2\eta_S\left(G_k,\Lambda_k\right) -\frac{5 G_k}{6\pi}\frac{(6 -\eta_h)}{(1-2\Lambda_{k})^2} \; .
\end{eqnarray}
Consistency of the truncation demands that all anomalous dimensions stay small. Thus the sign of $f_{\lambda}$ at the gravitational fixed point $(G_{\star},\Lambda_{\star})$ can be inferred by taking $\eta_h = 0$. The algebraic relation \eqref{eq:anomdimflow} for $\eta_S$ has the following solution:
\begin{equation}
\eta_S = \frac{16 G_k (1- \Lambda_k )}{(1 -2 \Lambda_k) (G_k - 12 \pi  (1 - 2 \Lambda_k ))}   \; ,  
\end{equation}
leading to:
\begin{equation}
    f_{\lambda} = -\frac{G_k \left(4 \pi  \left(-7 + 6 \Lambda_k+ 16 \Lambda^2_k\right)+5G_k \right)}{\pi(1-2 \Lambda_k)^2 (G_k -  12 \pi  (1 - 2 \Lambda_k))} \; .
\end{equation}
At leading order in $G_k$ and $\Lambda_k$,  in the physical regime $G_k>0$, we obtain $f_{\lambda}<0$, in agreement with previous results in the literature. This is a non-trivial cross check because we have worked with a different set of gauge parameters and, while physical results should be independent of such choices, truncations can introduce spurious gauge dependencies.
We note, however that $f_{\lambda}$ changes sign when:
\begin{eqnarray}
    \Lambda_{\star} < -\frac{3}{16} -\frac{\sqrt{121 \pi -20 G_{\star}}}{16 \sqrt{\pi }} \equiv \Lambda_{\text{crit}} \;.
\end{eqnarray}  
For $G_{\star} \ll 1$ we obtain $\Lambda_{\text{crit}} \approx -\frac{7}{8}$, and $\Lambda_{\text{crit}}$ increases with $G_{\star}$. This sign flip can be traced back to the $\eta_S$ term dominating in Eq.~\eqref{eq:flambdaapprox} and
we tentatively interpret this behavior as an artefact of our choice of gauge, because $\eta_S$ vanishes for $\beta=\alpha=0$ in $d=4$.  
We plan to come back to this point and perform a thorough study of the gauge-dependence of $\eta_S$ in future work.

\subsection{Universality of gravitational effects and absence of gravitationally-induced symmetry breaking}

The above calculation provides a concrete example of the universality of the gravitational contribution with respect to the internal symmetry of the matter because
$f_{\lambda}$ does not depend on $\lambda^{\bf abcd}$. Of course, universality in the sense of independence of internal symmetries does not mean scheme-independence, i.e.
$f_{\lambda}$ is RG-scheme-dependent.
    
The structure of the gravitational contribution implies that global symmetries of the matter sector are preserved. This is because the gravitational contribution to 
$\beta_{\lambda^{\bf abcd}}$ is linear in $\lambda^{\bf abcd}$ and thereby reproduces the internal symmetries encoded in the tensor structure of the coupling. More generally, if the regulator does not break a (non-anomalous) global symmetry, this symmetry will be  
 preserved by the flow. In all calculations in asymptotic safety to date, this statement also holds in the presence of gravity \cite{Eichhorn:2022gku}, see also \cite{Laporte:2021kyp}, and contradicts the so called "swampland conjecture" (reviewed, e.g., in \cite{Agmon:2022thq}) which, based on semi-classical arguments, states that global symmetries are broken by quantum gravity. Indeed, the swampland conjecture may not hold in asymptotic safety, see \cite{Eichhorn:2024rkc,Eichhorn:2022gku} for a detailed discussion of this point presenting several arguments in this direction as well as some caveats. 

Beyond the universal contribution $f_{\lambda}$, additional (non-universal) gravitational contributions to $\beta_{\lambda^{\bf abcd}}$ which are not accounted for in our truncation can exist and depend on the tensor structure of $\lambda^{\bf abcd}$. For instance, including a non-vanishing mass below brings a gravitational contribution to the flow of the double-trace coupling only, but not of the other two independent quartic couplings.

As a final point, $f_{\lambda}$ depends on additional couplings of the gravitational sector, see e.g. \cite{DeBrito:2019gdd}, but our truncation only accounts for the two most relevant ones. 

\section{The large-\texorpdfstring{$N$}{N} limit}
\begin{figure}[!t]
        \includegraphics[width=0.3\linewidth]{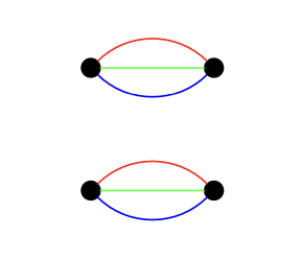} \quad
        \includegraphics[width=0.3\linewidth]{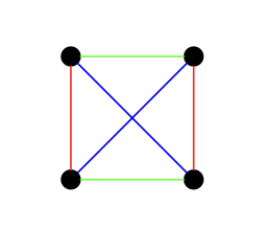}\\
        \includegraphics[width=\linewidth]{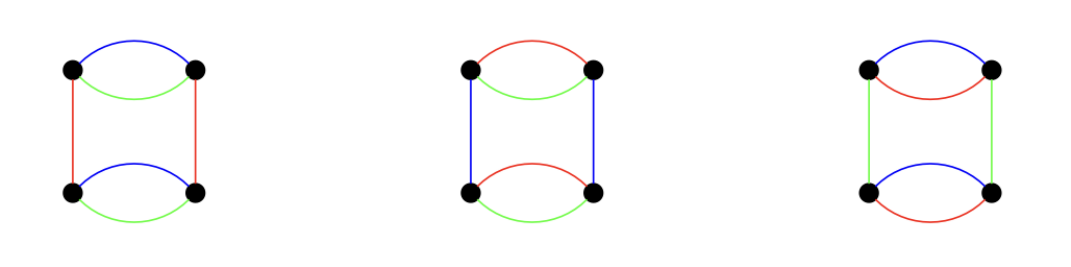}
    \caption{\label{fig:coloredgraphs} The 3-colored graphs representing the $O(N)^3$ quartic trace-invariants: the black dots represent the rank-3 tensor fields $\phi_{a_1a_2a_3}$ and the colored edges represent contractions between indices in the same position, red for $1$, green for $2$ and blue for $3$ (double trace and tetrahedron on the top row and the three pillow patterns at the bottom).}
\end{figure}

We now focus on the bosonic $O(N)^3$ tensor field theory in 4 dimensions where the coupling $\lambda^{\textbf{abcd}}$ is decomposed in a basis of three trace-invariants \cite{GurauBook}:
\begin{eqnarray}
    \lambda_k^{\textbf{abcd}} = \lambda_k^d \delta^d_{\textbf{abcd}} + \lambda_k^p \delta^p_{\textbf{abcd}} + \lambda_k^t \delta^t_{\textbf{abcd}} \label{eq:couplingsoriginal} \;,
\end{eqnarray}
where the  $O(N)^3$ invariant contraction patterns of indices are
\begin{eqnarray}
    \delta^d_{\textbf{ab;cd}}&=\delta_{a_1b_1}\delta_{a_2b_2}\delta_{a_3b_3}\delta_{c_1d_1}\delta_{c_2d_2}\delta_{c_3kd_3},\label{eq:doubletrace}\\
    \delta^p_{\textbf{ab;cd}}&=\frac{1}{3}\sum_{i=1}^3\delta_{a_ic_i}\delta_{b_id_i}\prod_{i\neq j}\delta_{a_jb_j}\delta_{c_jd_j},\label{eq:pillow}\\ 
    \delta^t_{\textbf{abcd}}&= \delta_{a_1b_1}\delta_{c_1d_1}\delta_{a_2c_2}\delta_{b_2d_2}\delta_{a_3d_3}\delta_{b_3kc_3} \label{eq:tetrahedron} \;. 
\end{eqnarray}
These are called the double trace, pillow and tetrahedral invariants and are depicted as 3-colored graphs in Fig.~\ref{fig:coloredgraphs}. Substituting this into  Eq.~\eqref{eq:quarticflow}, we obtain the 1-loop beta functions for the three corresponding couplings in the presence of gravity.

In the absence of gravity, the $O(N)^3$ tensor field theory admits a well defined large-$N$-limit for the 't Hooft couplings:
\begin{eqnarray}
    & \lambda^{d} \rightarrow \frac{\lambda^d}{N^3}  \; , \;\lambda^p \rightarrow \frac{ \lambda^p }{N^2}\;  , \;\lambda^t \rightarrow \frac{\lambda^t }{N^{\frac{3}{2}}}  \;,
\end{eqnarray}
or equivalently for the rescaled operators:
\begin{equation}
     \hat{\delta}^{d} = \frac{1}{N^3} \delta^{d} \; , \;\hat{\delta}^p = \frac{1}{N^2} \delta^p \;  , \;\hat{\delta}^t = \frac{1}{N^{\frac{3}{2}}} \delta^t \label{eq:operatorscalings}.
\end{equation}

The stability of the scalar potential is somewhat subtle to see in the basis in Eq.~\eqref{eq:couplingsoriginal}. It is appropriate to change basis in the space of invariants \cite{Benedetti:2019eyl,Berges:2023rqa}
and reparametrize the theory as:
\begin{eqnarray}
    \lambda^{\textbf{abcd}} = g \hat{\mathcal{P}}^{(t)} + g_1 \hat{\mathcal{P}}^{(1)} + g_2 \hat{\mathcal{P}}^{(2)} \;, \label{eq:orthonormalparam}
\end{eqnarray}
where:
\begin{equation}
    \hat{\mathcal{P}}^{(t)} \equiv \hat{\delta}^t\;\;,\;\; \hat{\mathcal{P}}^{(1)} \equiv 3(\hat{\delta}^p - \hat{\delta}^{d}) \;\;,\;\; \hat{\mathcal{P}}^{(2)} \equiv \hat{\delta}^d \;,
\end{equation}
and the associated couplings in the new basis are:
\begin{eqnarray}
    g \equiv \lambda^t \;,\; g_1 = \frac{\lambda^p}{3} \;,\; g_2 = \lambda^d +\lambda^p \label{eq:couplingredefinition} \;.
\end{eqnarray}
Denoting $ \hat{\mathcal{P}}^{(i)}\circ\hat{\mathcal{P}}^{(j)} = \sum_{\bf e,f}\hat{\mathcal{P}}^{(i)}_{\textbf{abef}}\hat{\mathcal{P}}^{(j)}_{\textbf{efcd}} $,we have at leading order at large $N$:\footnote{The product with one operator $\hat{\mathcal{P}}^{(t)}$ is understood as the symmetrization over all possible contraction channels.
}
\begin{align}\label{eq:products1}
    &  \hat{\mathcal{P}}^{(1)} \circ \hat{\mathcal{P}}^{(2)} = 0 \;\; , \;\; \hat{\mathcal{P}}^{(t)} \circ \hat{\mathcal{P}}^{(t)} =\frac{1}{9} \hat{\mathcal{P}}^{(1)} + \frac{1}{3}\hat{\mathcal{P}}^{(2)}
    \;, \nonumber \\
    & \hat{\mathcal{P}}^{(1)}\circ\hat{\mathcal{P}}^{(1)} = \hat{\mathcal{P}}^{(1)} \;,\;\hat{\mathcal{P}}^{(2)}\circ\hat{\mathcal{P}}^{(2)} = \hat{\mathcal{P}}^{(2)} \;,\nonumber \\
    & \hat{\mathcal{P}}^{(t)} \circ \hat{\mathcal{P}}^{(1,2)} \propto \frac{2}{3 N^{1/2}}\hat{\mathcal{P}}^{(1,2)} \;
    \end{align}
This leads
to a particular case of the one-loop algebras of multi-scalar models \cite{Flodgren:2023tri}. Seen as operators from $ \textbf{cd} $ to $\textbf{ab}$, $ \hat{\mathcal{P}}^{(1,2)}_{\textbf{abcd} } $ are a couple of (mutually) orthogonal projectors, therefore the corresponding interaction quartic monomials are explicitly positive. These two terms are stable if $g_1,g_2>0$ which translates into 
$\lambda^p > 0 $ and $\lambda^d > - \lambda^p$, yielding a necessary condition for stability of the theory when neglecting the higher-order terms in the scalar potential.

However, the remaining tetrahedral interaction monomial does not have a definite sign, i.e. depending on the configuration of the tensor field, either sign may be realized; therefore one cannot immediately conclude on the stability of the theory if all the couplings are real. 

\subsection{The large-\texorpdfstring{$N$}{N} limit without gravity}

The RG flow of the $O(N)^3$ tensor field theory at two loops exhibits a non-unitary Wilson-Fisher like fixed point in $d=4-\epsilon$ dimensions \cite{Giombi:2017dtl} and, if one insists on all the couplings being real, the usual Landau pole in $d=4$.

Inspired by the long-range version of the model \cite{Benedetti:2019eyl}, in \cite{Berges:2023rqa} 
the authors considered the $O(N)^3$ theory in $d=4$, but with a \emph{purely imaginary} tetrahedral coupling:
\begin{eqnarray}
 \lambda_k^t \rightarrow i \lambda_k^t \label{eq:imaginarymapping} \; ,
\end{eqnarray}
while keeping $g_1$ and $g_2$ real. While having an imaginary coupling would seem to doom the model to be unstable and non-unitary, it turns out that this version of the model is stable (the real part of the bare action is bounded from below), has a real quantum effective action at leading large-$N$ order non-perturbatively in the coupling constant and  is asymptotically free in the large-$N$ limit \cite{Berges:2023rqa}. 

\begin{figure*}[!t]
        \includegraphics[width=0.45\linewidth]{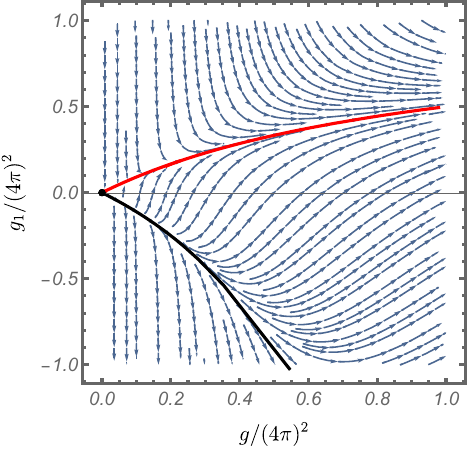}
        \quad
        \includegraphics[width=0.45\linewidth]{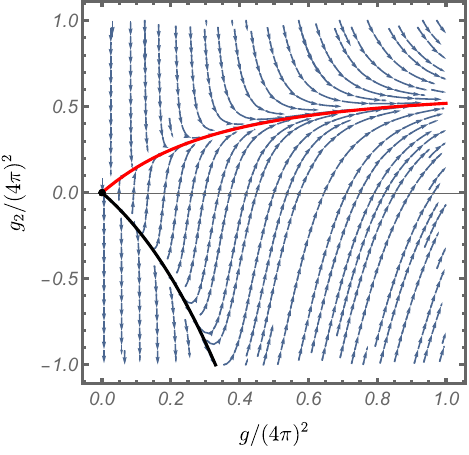}
    \caption{\label{fig:nogravFLOW} The 2-loop RG-flow chart of the couplings for imaginary tetrahedral coupling. The arrows point towards the infrared $k\to 0$.}
\end{figure*}

Asymptotic freedom occurs because switching to an imaginary coupling as in \eqref{eq:imaginarymapping} changes the nature of fluctuations of the scalar field from screening to antiscreening. At one loop,  Eqs.~\eqref{eq:products1} imply that mixed terms $ g_1 \cdot g_2$, $g\cdot g_1$ and $g\cdot g_2$ cannot occur, but $g$ back reacts on the flow of $g_1$ and $g_2$. This signals the possible existence of enhanced symmetries and indeed, for vanishing tetrahedral and pillow couplings the model exhibits an enhanced $O(N^3)$ global symmetry which is preserved by the RG. 
At 2-loops, with complex tetrahedral coupling, the beta functions of the model read
\begin{align}\label{eq:systemnograv1}
    &\partial_t g = - 2 \frac{g^3}{(4 \pi)^4}  \; , \nonumber \\
    &\partial_t g_1 = 2\left(\frac{g_1^2 - g^2}{(4\pi)^2} + \frac{g^2 g_1}{(4\pi)^4}\right) \;,  \nonumber \\
    &\partial_t g_2 = 2\left(\frac{g_2^2 - 3g^2}{(4\pi)^2} + 5\frac{g^2 g_2}{(4\pi)^4}\right) \; .
\end{align}
The flow of the tetrahedral coupling is decoupled from the other two at all orders. This system of flow equations leads to the RG flow depicted in Fig.~\ref{fig:nogravFLOW}. 

Eq.~\eqref{eq:systemnograv1} shows that $g$ is an asymptotically free coupling. For $g_1$ and $g_2$, the situation is more subtle. At $g=0$, the Gaussian fixed point is infrared attractive (resp.~repulsive) in $g_{1,2}$ for positive (resp. negative) values of $g_{1,2}$, because it is a double zero of the beta function. In contrast, at $g \neq 0$, both $\beta_{g_{1,2} }$ are negative at $g_{1,2}=0$, i.e., both $g_{1,2}$ grow towards the IR when starting from $g_{1,2} =0,\, g >0$. It follows that there exist asymptotically free RG trajectories in the space of couplings. The precise situation is depicted in Fig.~\eqref{fig:nogravFLOW}, where two separatrices (in red and black in the figure) are found. Above the red separatrix, the flow is not UV complete, while below the black separatrix, the flow is UV complete, but the IR values of $g_{1,2}$ are negative.
 Within the wedge between these separatrices, the theory flows from a weakly coupled UV regime to a strongly coupled one in the IR such that $g_{1,2}$ are positive and the theory is stable.\footnote{As the flow is
symmetric under $g \rightarrow -g$, the structure of the flow is even with respect to $g=0$.}

The aim of this paper is to understand how the structure of the RG flow in the large-$N$ limit 
changes when gravitational back reaction on matter is taken into account.
A priori, there are four conceivable scenarios regarding the fate of the gravitational contribution in the large-$N$-limit:
\begin{enumerate}
\item[i)] The gravitational contribution to the matter beta functions is entirely suppressed and the matter sector decouples from gravity.
\item[ii)] A part of the gravitational contribution is suppressed, but another part of it contributes.
\item[iii)] The entire gravitational contribution comes at the same order as the large-$N$ matter self interaction contributions and at leading order the beta functions contain both.
\item[iv)] The gravitational contribution dominates the beta functions.
\end{enumerate}
Which of these scenarios is realized depends entirely on the scaling of the gravitational fixed-point values with $N$. While various results on the dependence of the fixed point on $N$ exist in the literature, \cite{Dona:2013qba,Meibohm:2015twa,Biemans:2017zca,Christiansen:2017cxa,Alkofer:2018fxj,Wetterich:2019zdo,Sen:2021ffc, Korver:2024sam}, and also \cite{Narain:2009fy, Dona:2013qba, Percacci:2015wwa, Labus:2015ska, Meibohm:2015twa, Dona:2015tnf, Biemans:2017zca, Alkofer:2018fxj, Eichhorn:2018akn,Wetterich:2019zdo, Laporte:2021kyp, Sen:2021ffc,deBrito:2021pyi,deBrito:2023myf} specifically for scalar fields, this question is not yet settled.
In small truncations for the gravity sector, such as the one we have considered so far, the fixed-point value $G_*$ grows with $N$, and no real fixed point $(G_*,\Lambda_*)$ exists beyond a critical value $N_{\rm crit}$. On the other hand, it has been argued that within such a truncation the effect of matter on 
$G_*$ can be absorbed for any $N$ into a renormalization, and the fixed point always exists \cite{Christiansen:2017cxa}. Moreover, technical choices such as the application of the background field approximation versus the fluctuation-approach, seem to result in qualitative differences in the large-$N$ behavior, see \cite{Eichhorn:2018akn} for a careful analysis.
To make matters worse, the large-$N$-scaling of $G_{\ast}$ does not in fact suffice in order to conclude on the fate of the gravitational contributions, as the latter are also suppressed (enhanced) for $\Lambda_{\ast} \rightarrow -\infty$ ($\Lambda_{\ast} \rightarrow 1/2$) and much of the uncertainty around the large-$N$ limit
comes precisely from the non-trivial interplay between matter fluctuations and the full, nonperturbative structure of the propagator of metric fluctuations.
In summary, it is not yet  robustly established whether or not the gravitational fixed point exists in the large-$N$ limit and how it scales with $N$. 

From here on we will resort to a more phenomenological approach. 
Of the four scenarios listed above, ii) and iii) are clearly the most interesting ones. We therefore make the assumption that in the large-$N$ limit $f_{\lambda}$ survives and that additional gravitational contributions to the matter beta functions can in a first approximation be neglected. 

%
%
\subsection{
Large-\texorpdfstring{$N$}{N} limit in the presence of gravity \texorpdfstring{$f_{\lambda}<0$}{flambda<0}}

We posit the large-$N$ beta functions:

\begin{align}\label{eq:systemgrav}
    &\partial_t g =  -f_{\lambda} g - 2 \frac{g^3}{(4 \pi)^4}  \nonumber,\\
    &\partial_t g_1 =  -f_{\lambda} g_1 + 2\left(\frac{g_1^2 - g^2}{(4\pi)^2} + \frac{g^2 g_1}{(4\pi)^4}\right),\nonumber \\ 
    &\partial_t g_2 =  -f_{\lambda} g_2 + 2\left(\frac{g_2^2 - 3g^2}{(4\pi)^2} + 5\frac{g^2 g_2}{(4\pi)^4}\right)\; , 
\end{align}
where according to our previous discussion, the gravitational contribution is blind to the details of the internal symmetry of the matter couplings, i.e., all the gravitational contributions come with the same coefficient $-f_{\lambda}$.

The inclusion of gravity has several effects. As the gravitational contribution is screening ($f_{\lambda}<0$), $g$ is no longer asymptotically free and the Gaussian fixed point becomes infrared attractive in all 
three couplings. Moreover, because the gravitational contribution is linear in the quartic couplings, the Gaussian fixed point is no longer degenerate and it splits into four fixed points, listed together with their critical exponents (defined as minus the eigenvalues of the stability matrix) in Tab.~\ref{table:vanishingtetrahedral}.
As the fixed-point values of $g_1$ and/or $g_2$ are negative, none of these fixed points is deemed viable.
\begin{table}[ht]
    \[
    \renewcommand{\arraystretch}{1.5} 
    \setlength{\tabcolsep}{15pt}      
    \begin{array}{|c|c|c|c|c|c|}
    \hline
    g_{ \ast}/(4\pi)^2 & g_{1,\ast}/(4\pi)^2 & g_{2,\ast}/(4\pi)^2 & \theta_g & \theta_{g_1} & \theta_{g_2} \\
    \hline
    0 & 0 & 0 & f_\lambda & f_\lambda & f_\lambda \\
    \hline
    0 & \frac{f_\lambda}{2} & 0 & f_\lambda & -f_\lambda & f_\lambda \\
    \hline
    0 & 0 & \frac{f_\lambda}{2} & f_\lambda & f_\lambda & -f_\lambda \\
    \hline
    0 & \frac{f_\lambda}{2} & \frac{f_\lambda}{2} & f_\lambda & -f_\lambda & -f_\lambda \\
    \hline
    \end{array}
    \]
\caption{Fixed points with vanishing tetrahedral coupling. 
The stability matrix is diagonal in all the cases and the three operators in the bare action are the eigen-perturbations of the fixed points. 
}
\label{table:vanishingtetrahedral}
\end{table}

\begin{figure*}[!t] 
        \includegraphics[width = 0.45 \textwidth]{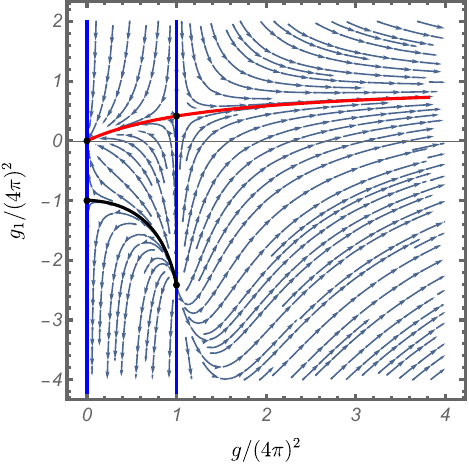}
        \includegraphics[width = 0.45 \textwidth]{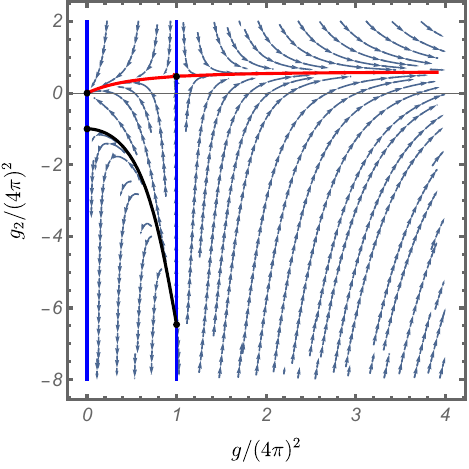}
    \caption{Flow trajectories of $g_1$ (left) and $g_2$ (right) in the presence of asymptotically safe gravity, as obtained from the system Eq.~\eqref{eq:systemgrav} for $f_{\lambda} = -2$ 
    (chosen purely for purposes of illustration), where arrows point in the direction of decreasing RG-scale $k$. The blue line at $g = (4\pi)^2$ is a phase transition line between an IR trivial and a strongly coupled theory.}
    \label{fig:gravFLOW} 
\end{figure*}

The competition between the screening gravity term $-f_{\lambda}g>0$ and the antiscreening matter term $-2 g^3<0$ in $\beta_g$ yields interacting fixed points with $g_* \neq 0$ with and at least one relevant direction. Specifically, we obtain four (eight if we account for the two distinct sign choices for $g_{*}$) interacting fixed points:
\begin{eqnarray}
     \frac{g_{\ast}}{(4\pi)^2} &=& \pm \sqrt{-\frac{f_\lambda}{2}} \; ,\nonumber\\ 
     \frac{g_{1, \ast}}{(4\pi)^2} &=& \frac{1}{2}\left(f_{\lambda} \pm \sqrt{f_{\lambda}^2 - 2f_{\lambda}}\right) \; ,\nonumber\\
     \frac{g_{2, \ast}}{(4\pi)^2} &=& \frac{1}{2}\left(3 f_{\lambda} \pm \sqrt{3}\sqrt{3f_{\lambda}^2 - 2 f_{\lambda}}\right) \; .
\end{eqnarray}
The fixed-point values $g_{*}$, $g_{1,*}$, $g_{2,*}$ are real for any $f_{\lambda}$, as long as gravity is screening ($f_{\lambda}<0$). At all these fixed points,
the stability matrix is diagonal in $g_{1,2}$ and the corresponding operators are eigen-perturbations with critical exponents $\theta_1$ and $\theta_2$ respectively. The remaining critical exponent, which we call $\tilde{\theta}$, is minus the remaining diagonal entry of the stability matrix and is associated to an eigen-perturbation which is a linear combination of the three operators. This is displayed in Fig.~\ref{fig:gravFLOW}: the interacting fixed points (at $g>0$) always have an eigen-direction parallel to the vertical axes. The corresponding critical exponents are:

\begin{eqnarray}\label{eq:expo}
    \tilde{\theta}& =& -2 f_{\lambda} > 0, \nonumber\\
    \theta_1 &=& \mp 2 \sqrt{-f_{\lambda}(2- f_{\lambda})},\nonumber\\
    \theta_2 &=& \mp 2 \sqrt{3} \sqrt{-f_{\lambda} (2 - 3 f_{\lambda})} \; .
    \end{eqnarray}
All these interacting fixed points have one relevant direction corresponding to $\tilde{\theta}>0$ and, as a function of the signs, can have up to two additional relevant directions if $\theta_1,\theta_2>0$. 

A priori all these fixed points constitute UV completions of the model with different amounts of predictive power, and are distinguished only by the signs of $g_{1,2}$. These signs are crucial for the stability of the potential and the boundedness from below of the 
fixed-point action.
This rules out the fixed points with negative $g_1$ or $g_2$ and leaves us with a single fixed point (or rather, two fixed points, related by $g\rightarrow -g$), which has one relevant direction. The resulting RG trajectories flow either towards the Gaussian fixed point or towards a strong-coupling regime in the IR along the two red separatrices, as depicted in Fig.~\eqref{fig:gravFLOW}.

From Fig.~\ref{fig:gravFLOW}, one can also see how the fixed point generates universality, as all the trajectories are pulled to the red separatrix, which is the IR stable manifold of the fixed point (strictly speaking the stable manifold has another direction corresponding to the mass).

\section{Conclusions and outlook}

In this paper, we found tentative hints for a first example of a theory of scalar matter together with gravity that is asymptotically safe at nonzero quartic (matter self interaction) coupling. This is in distinction to other works in the literature, where the quartic coupling vanishes at an asymptotically safe fixed point with gravity \cite{Narain:2009fy, Percacci:2015wwa, Eichhorn:2020sbo} and matter sectors, e.g. for dark-matter models, are often chosen with additional degrees of freedom that couple to the scalar and can generate a non-vanishing scalar fixed-point potential, see e.g. \cite{Reichert:2019car, Eichhorn:2019dhg, Eichhorn:2020kca, Eichhorn:2020sbo}.

Our qualitatively distinct result follows from the fact that, on its own, the pure-matter theory is asymptotically free. 
In the $O(N)^3$ case with imaginary tetrahedral coupling, we found that the presence of asymptotically safe gravity transforms some of the previously asymptotically free trajectories to asymptotically safe ones that still flow towards a strongly coupled regime in the IR. In particular, the gravitational contribution generates an RG-trajectory along which the theory is asymptotically safe, bounded and strongly coupled in the IR, the red separatrix in Fig.~\ref{fig:gravFLOW} (on the right of the transition line). 

The separatrix is also interesting in the context of \emph{effective asymptotic safety} \cite{Held:2020kze}. This is the idea that the UV completion of the theory is not asymptotically safe, such that the RG flow we have calculated is only a viable description below some transition scale $k_{\rm trans}$. The underlying microscopic theory, which need not even be a QFT \cite{deAlwis:2019aud,Basile:2021krr, Basile:2024oms}, provides initial conditions for the couplings in its effective QFT description at $k_{\rm trans}$. If these lie in the basin of attraction of the fixed point, the corresponding RG trajectories will approach the critical surface of the fixed point and may in the deep IR become indistinguishable from fixed-point trajectories for all practical purposes \cite{Percacci:2010af}. In our case,
while the trajectories above and below the red separatrices do not consist viable UV completions, they provide the same universal IR predictions. 

Our analysis relies on taking the number of scalar fields to infinity and, unfortunately,  the structure of the gravitational fixed point itself is not yet well-understood in this limit. We 
conjectured that the main gravitational contribution to the beta function of quartic couplings, which structurally resembles a scaling dimension, remains present in this limit.

There are several interesting directions for future work. First, it is interesting to understand the fixed-point structure including subleading corrections in $1/N$ or even extending to small values of $N$, where the impact of scalar fields on the gravity fixed point is already robustly understood and small. Second, achieving a better control of the gravitational fixed point at large $N$ and understanding whether the gravitational contribution in the beta function of quartic couplings persists is clearly an important open issue. Third, the fixed-point structure we found here may potentially be of phenomenological interest for hidden/dark sectors.

\section{Acknowledgements}
We thank Marc Schiffer and Hannes Kepler for discussions.
This work is funded by the Deutsche Forschungsgemeinschaft (DFG, German Research Foundation) under Germany’s Excellence Strategy EXC 2181/1 - 390900948 (the Heidelberg STRUCTURES Excellence Cluster).
----

\color{black}
\bibliography{references}

\end{document}